\documentclass{aa}

\usepackage{graphicx}
\usepackage{natbib}
\usepackage{txfonts}
\usepackage{color}

\def\arcmin{\hbox{$^\prime$}}
\def\arcsec{\hbox{$^{\prime\prime}$}}

\def\flux{erg s$^{-1}$ cm$^{-2}$}

\def\lum{erg s$^{-1}$}

\def\aap{A\&A}

\def\apjs{ApJS}

\def\apjl{ApJLett}

\def\smcx3{SMC\,X-3}

\newcommand {\be}{\begin {equation}}
\newcommand {\ee}{\end {equation}}

\begin{document}

   \title{SMC\,X-3: the closest ultraluminous X-ray source powered by a neutron star with non-dipole magnetic field}

   \author{S.~S.~Tsygankov \inst{1,2}
          \and  V. Doroshenko \inst{3}
          \and  A. A. Lutovinov \inst{2,4}
          \and A. A. Mushtukov \inst{5,6,2}
          \and J. Poutanen \inst{1,7}
          }

   \institute{Tuorla Observatory, Department of Physics and Astronomy,  University of Turku,
              V\"ais\"al\"antie 20, 21500 Piikki\"o, Finland
              \email{sergey.tsygankov@utu.fi}
       \and
             Space Research Institute of the Russian Academy of Sciences, Profsoyuznaya Str. 84/32, Moscow 117997, Russia
       \and
              Institut f\"ur Astronomie und Astrophysik, Universit\"at T\"ubingen, Sand 1, D-72076 T\"ubingen, Germany
       \and
             Moscow Institute of Physics and Technology, Moscow region, Dolgoprudnyi, Russia
       \and
            Anton Pannekoek Institute, University of Amsterdam, Science Park 904, 1098 XH Amsterdam, The Netherlands
       \and
           Pulkovo Observatory of the Russian Academy of Sciences, Saint Petersburg 196140, Russia
       \and
          Nordita, KTH Royal Institute of Technology and Stockholm University, Roslagstullsbacken 23, SE-10691 Stockholm, Sweden
       \
          }
   \titlerunning{\smcx3: the closest  ULX-pulsar}
   \authorrunning{S. Tsygankov et al. }
   \date{Received; accepted}


  \abstract
  {}
   {Magnetic field of accreting neutron stars determines their overall behavior including the maximum possible luminosity. Some models require an above-average magnetic field strength ($\gtrsim 10^{13}$~G) in order to explain super-Eddington mass accretion rate in the recently discovered class of pulsating ultraluminous X-ray sources (ULX). The peak luminosity of \smcx3\ during its major outburst in 2016--2017 reached $\sim2.5\times10^{39}$~\lum comparable to that in ULXs thus making this source the nearest  ULX-pulsar. Determination of the magnetic field of \smcx3\ is the main goal of this paper. }
   {\smcx3\ belongs to the class of transient X-ray pulsars with Be optical companions, and exhibited a   giant outburst in July 2016 -- March 2017. The source has been observed during the entire outburst with the {\it Swift}/XRT and {\it Fermi}/GBM telescopes, as well as the {\it NuSTAR} observatory. Collected data allowed us to estimate the magnetic field strength of the neutron star in \smcx3\ using several independent methods.}
{Spin evolution of the source during and between the outbursts
and the luminosity of the transition to so-called propeller regime in the range of $(0.3-7)\times10^{35}$~erg~s$^{-1}$ imply relatively weak dipole field of $(1-5)\times10^{12}$~G. On the other hand, there is also evidence for much stronger field in the immediate vicinity of the neutron star surface. In particular, transition from super- to sub-critical accretion regime associated with cease of the accretion column and very high peak luminosity favor an order of magnitude stronger field. This discrepancy makes \smcx3\ a  good candidate to posses significant non-dipolar components of the field, and an intermediate source between classical X-ray pulsars and accreting magnetars which may constitute an appreciable fraction of ULX population.}
   {}


   \maketitle

%

\section{Introduction}

Magnetic field of a neutron star (NS) defines observational properties for a
broad range of systems. Strongly magnetized accreting NSs, or X-ray pulsars (XRPs), are among the most prominent NS systems and are
being actively studied, particularly in connection with the possibility of
super-Eddington accretion. The main source of information about the strength of
magnetic fields in XRPs is associated with the so-called cyclotron scattering resonance absorption features (CSRF)
observed in the energy spectra of some sources. Unfortunately, this method is
restricted by a sensitivity and energy range of X-ray telescopes and relatively
soft spectra of XRPs. This implies a very limited number of XRPs with known
magnetic fields, all of which fall into a narrow range
$B\sim(1-8)\times10^{12}$~G \citep[see review by][]{2015A&ARv..23....2W}
defined by aforementioned selection effects. At the same time magnetic fields of
very bright XRPs (including the pulsating ultraluminous X-ray sources, ULXs) are
expected to be $\gtrsim 10^{13}$~G \citep{2015MNRAS.454.2539M}, and it would be
very important to find sources with similar fields among the ordinary accreting
XRPs.

Fortunately, besides the spectroscopy, pulsars timing properties can be used to
independently estimate the magnetic field strength. Accretion torque affecting
the NS depend on the magnetosphere size and, although model-dependent
\citep[see, e.g.,][]{2016ApJ...822...33P}, can be used to estimate the field.

Detection of a centrifugal inhibition of accretion, known as the propeller
effect \citep{1975A&A....39..185I,1986ApJ...308..669S} can also be used to
estimate the magnetosphere size and thus the magnetic field. The reliability of
this approach has been recently demonstrated by
\citet{2016A&A...593A..16T,2016MNRAS.457.1101T} and \citet{2017ApJ...834..209L}, who showed
that the magnetic field values estimated from the propeller effect are in good
agreement with measurements from independent methods (including the direct
estimate based on the observed CSRF energies).

\smcx3\ was discovered with the {\it SAS-3} observatory as a bright
source in the Small Magellanic Cloud by
\citet{1978ApJ...221L..37C}. The source was reported to have a
luminosity of $7\times10^{37}$\,\lum\ in the $2-11$ keV energy band,
and a relatively hard spectrum (with photon spectral index of
$\simeq 1$). Despite of the early proposed optical identification
\citep{1977IAUC.3134....3V,1978ApJ...223L..79C} the nature of the
source remained uncertain for a long time. Only in 2004,
\citet{2004MNRAS.353.1286E} using the {\it Chandra} data had shown
that the  7.78~s pulsar found by \citet{2004AIPC..714..337C} in the
{\it RXTE} data can be identified with \smcx3. Based on the
precise X-ray position reported by \citet{2004MNRAS.353.1286E},
\citet{2004AJ....128..709C} studied the long term optical light curve
of \smcx3\ and established its counterpart as a O9e star. A number of
outbursts detected with the {\it RXTE} observatory in 1999--2009
allowed to find a periodicity in the source activity, likely
associated with the orbital motion with the period of $\simeq45$ days
\citep{2004AIPC..714..337C}. Based on the long-term spin period evolution
observed by the {\it RXTE}, \citet{2014MNRAS.437.3863K} estimated the expected
magnetic field of the NS in \smcx3\ as
$B\simeq2.9\times10^{12}$\,G.  We note that in this case a CSRF
in the source spectrum at the energy of $\sim26$ keV
should be observable (accounting for the gravitational redshift). Another estimate of the
magnetic field strength $B\simeq7.3\times10^{12}$\,G was obtained
recently by \cite{2017arXiv170102983W} based on the assumption that
the source was spinning close to an equilibrium in the tail of the
2016--2017 outburst.

Current 2016--2017 outburst from \smcx3\ was detected with the {\it MAXI} monitor and was
initially designated as a possible new source MAXI\,J0058-721
\citep{2016ATel.9348....1N}. The follow-up observations with the {\it
Swift}/XRT telescope allowed to establish, however, that the new transient is
in fact a known source \citep{2016ATel.9362....1K}, and to perform a timing and
spectroscopy of this object in soft X-rays \citep{2016ATel.9370....1K}.
Moreover, follow-up observations with the {\it NuSTAR} observatory allowed to
investigate the broadband spectrum of \smcx3\ for the first time. It was shown
that the \smcx3\ spectrum in the 3-50 keV energy band can be well described by
an exponentially cutoff power law model with the photon index of
$\Gamma\simeq0.5$ and the folding energy of $E_{\rm fold}\simeq12$ keV.
Additionally a black body component with the temperature of $kT\simeq1.78$ keV
and an emission line from neutral iron with equivalent width of 70 eV were
observed in the spectrum \citep{2016ATel.9404....1P}. It is important to
emphasize that no other obvious spectral features, including the CSRF, were reported \citep{2016ATel.9404....1P}.

Extremely high bolometric luminosity of the source $L_{\rm
peak}\sim2.5\times10^{39}$~\lum, reached during this outburst, makes \smcx3\
unique among transient X-ray pulsars with Be optical companions (Be/XRPs). In
fact, due to the high luminosity it can be formally attributed to the group of
ULXs. Its proximity to us would make it the
closest ULX in this case, and furthermore, the closest  ULX-pulsar. In this
work we present results of the monitoring program performed with the {\it
Swift}/XRT and {\it Fermi}/GBM telescopes, as well as the {\it NuSTAR}
observatory during recent outburst in July 2016 -- March 2017. These data
allowed us to estimate the magnetic field strength of the NS in the
system using several independent methods and to conclude that the source likely
has a non-dipole configuration of the magnetic field.

\section{Observations}

\subsection{{\it Swift}/XRT data}

The best facilities for long-term monitoring programs in a broad
range of fluxes are currently provided by the {\it Swift} observatory
\citep{2004ApJ...611.1005G}. During the current 2016--2017 outburst regular
observations of \smcx3\ were performed with the XRT telescope
\citep{2005SSRv..120..165B} in the soft X-ray band (0.5--10 keV) providing
both high sensitivity and flexibility. The data analyzed in this work
were collected between MJD 57611 and MJD 57840.

The XRT telescope observed \smcx3\ in both Windowed Timing (WT;
providing good temporal resolution) and Photon Counting (PC)
modes. Final scientific products (spectrum in each observation) were
produced using online tools provided by the UK Swift Science Data
Centre \citep{2009MNRAS.397.1177E}.\footnote{\url{http://www.swift.ac.uk/user_objects/}}

The spectra were grouped to have at least 1 count per bin and fitted
using a simple power law model modified with the photoelectric
absorption ({\sc phabs$\times$powerlaw} model in the {\sc xspec} package) and
Cash statistic \citep{1979ApJ...228..939C}. To avoid any problems
caused by the calibration uncertainties at low
energies,\footnote{\url{http://www.swift.ac.uk/analysis/xrt/digest_cal.php}}
we restricted the spectral analysis to the 0.7--10 keV and 0.5--10 keV energy
bands for the data in WT and PC modes, respectively.

We found that at all luminosities the spectra of \smcx3\ can be well
described with a simple absorbed power-law model with the photon index being in
the range of 0.5--1.1. The spectral analysis did not reveal any significant
absorption in excess of the Galactic hydrogen column density measured in this
direction $N_{\rm H}=0.066\times10^{22}$ cm$^{-2}$ \citep{1990ARA&A..28..215D}.
To make the spectral approximation more robust we fixed the $N_{\rm H}$ value at
this value. \cite{2017arXiv170102983W} claimed an appearance of the black-body
component at very soft energies which were ignored in our analysis.

\begin{figure}
\centering
\includegraphics[width=0.98\columnwidth, bb=55 265 575 680]{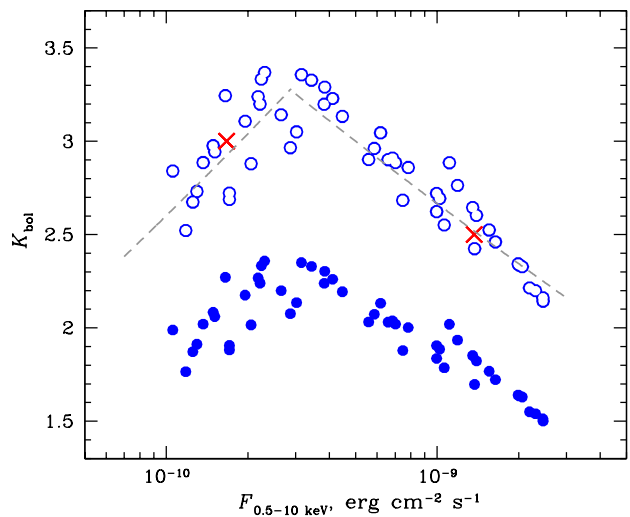}
\caption{The bolometric correction factor as a function of the source
  intensity (open circles). Filled circles show the dependence of the
  ratio of the total flux in $0.5-10$ keV plus $15-50$ keV to the
  $0.5-10$ keV flux as a tracer of the bolometric correction factor
  ($K_{\rm bol}$).  Red crosses show the $K_{\rm bol}$ value and the
  source flux (in $0.5-10$ keV range) from the broad-band spectra
  collected with the {\it NuSTAR} observatory. }\label{fig:bolom}
\end{figure}

\subsection{{\it NuSTAR} data}

The {\it NuSTAR} observatory consists of two co-aligned identical X-ray
telescope systems (FPMA and FPMB) operating in a wide energy range from 3 to
79 keV \citep{2013ApJ...770..103H}. Thanks to the unique multilayered
mirrors, {\it NuSTAR} has an unprecedented sensitivity in hard X-rays ($>10$
keV) and is ideally suited for the broadband spectroscopy of different
objects, including X-ray pulsars, and searching for the CSRFs in their
spectra.

\smcx3\ has been observed with {\it NuSTAR} twice during the current
2016--2017 outburst (ObsIDs 90201035002 and 90201041002) with the aim
of measuring its hard X-ray spectrum. Preliminary results of the
analysis of the first observation with the bolometric luminosity
$\sim10.2\times10^{38}$\,\lum\ were discussed above
\citep{2016ATel.9404....1P}. The second observation was performed at
our request three months later on MJD 57704.8 when the source
luminosity was $\sim1.9\times10^{38}$\,\lum\ that is about an order of
magnitude lower in comparison to the first observation.

The raw data obtained during both observations were processed to
produce cleaned event files for the FPMA and FPMB modules using the
standard {\it NuSTAR} Data Analysis Software ({\sc NuSTARDAS}) v1.7.1
provided under {\sc HEASOFT v6.20} with the CALDB version
20170222. Using the {\sc nuproducts} routine, we extracted the source
spectra from the circlar region with radius of 120\arcsec. The
background spectrum was extracted in the region of the same radius
located 5\arcmin\ from the source position.

\subsection{Bolometric correction}

For any meaningful discussion of the
observed source properties a bolometric correction has to be estimated for the
observed flux in soft energy band. To do that we used two available {\it
NuSTAR} observations as reference points and the dependence of the ratio of the
total source flux in the $0.5-10$ keV plus $15-50$ keV energy bands to the
source flux in the $0.5-10$ keV energy band. The flux in the $15-50$ keV band
has been estimated using {\it Swift}/BAT transient monitor\footnote{\url{http://swift.gsfc.nasa.gov/results/transients/}}
light curve of the source. The
ratio $(F_{\rm 0.5-10 keV}+F_{\rm 15-50 keV})/F_{\rm 0.5-10 keV}$ as a function
of $F_{\rm 0.5-10 keV}$ is shown in Fig.~\ref{fig:bolom} with filled circles.
To convert this ratio to the bolometric correction factor $K_{\rm bol}$ we
rescaled it to match the values calculated from the spectral parameters
obtained from two {\it NuSTAR} observations (marked by red crosses). The result
of this adjustment shown with the open circles was fitted with a broken linear
model (shown with the grey dashed line in the same figure). We estimate that
residual systematic uncertainty in the correction factor does not exceed 5\%.
Due to unknown shape of the source broadband spectrum at low fluxes we fixed
the bolometric correction factor value at 2 for all observations with $F_{\rm
0.5-10 keV}\textless4\times10^{-11}$~\flux. In the following analysis we apply
this correction to all observational data and refer to the bolometrically
corrected fluxes and luminosities, unless stated otherwise.

\section{Results}
\label{sec:res}

\begin{figure}
\centering
\includegraphics[width=0.98\columnwidth, bb=20 155 575 680]{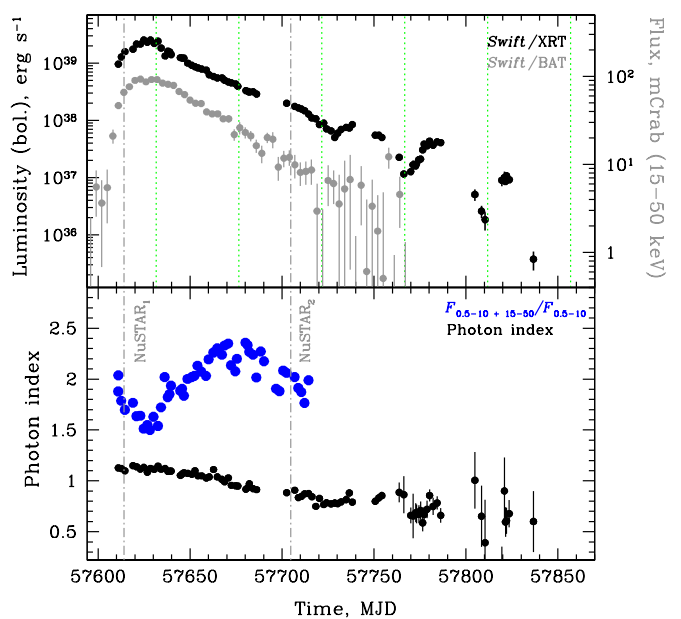}
\caption{{\it Upper panel}: The bolometric light curve of
  \smcx3\ obtained with the {\it Swift}/XRT telescope (black
  points). Luminosity is calculated from the unabsorbed flux under
  assumption of the distance to the source $d=62$ kpc and bolometric
  correction factors from Fig.~\ref{fig:bolom}. Grey dots represent
  flux in the 15--50 keV band from the {\it Swift}/BAT monitor (in the
  units of mCrab, right axis). Green vertical dotted lines correspond to the times
  of the periastron passages. {\it Bottom panel}: Evolution of the
  photon index and ratio of fluxes $(F_{\rm 0.5-10 keV}+F_{\rm 15-50
    keV})/F_{\rm 0.5-10 keV}$ over the outburst are shown with black
  and blue points, respectively. Vertical dash-dotted lines correspond
  to the times of the {\it NuSTAR} observations. }\label{fig:lc}
\end{figure}

The light curve of \smcx3\ observed with the {\it Swift}/XRT telescope
is shown in the upper panel of Fig.~\ref{fig:lc} with black
points. The source flux was estimated using the bolometric and
absorption corrections described above and assuming a distance to the
source of 62 kpc \citep{2012AJ....144..107H}. The observed light curve
looks rather complicated.  One can see from Fig.~\ref{fig:lc} a clear
transition to a faster luminosity decay after MJD $\sim$57710. This
transition can be understood in terms of the thermal-viscous
instability model as a moment when temperature at the outer radius of
the accretion disk reached the critical temperature of $\sim 6500$~K
causing the decrease of the local viscosity and corresponding a
decline of the mass accretion rate onto the compact object \citep[see,
  e.g., ][]{2001NewAR..45..449L}. This results in the fast decay of
the luminosity observed, particularly, from XRPs in the very end of
their outbursts \citep[see recent works by
][]{2016A&A...593A..16T,2017ApJ...834..209L}. As can be seen from
Fig.~\ref{fig:lc} this decay is not smooth and was interrupted with
temporal re-brightenings a few times on MJD $\sim$57730, MJD
$\sim$57770 and MJD $\sim$57810, near the corresponding periastron
passages.

\subsection{Orbital parameters}
\label{sec:opars}

\begin{figure}[t]
    \centering
        \includegraphics[width=\columnwidth]{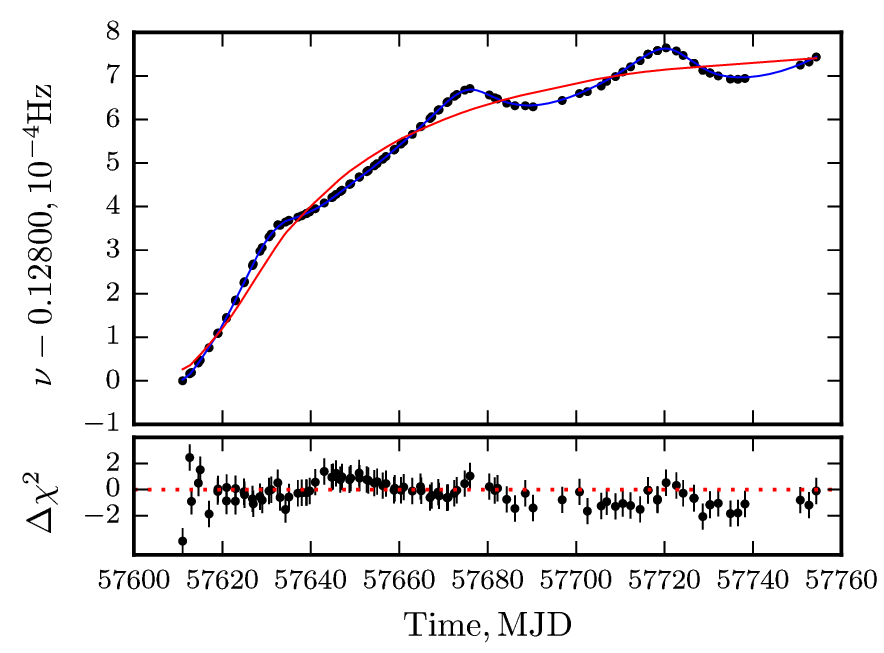}
    \caption{Spin evolution of \smcx3\ as observed by {\it Fermi}/GBM
      and {\it Swift}/BAT (black points), and the best-fit models for
      intrinsic (red) and observed (blue) pulse frequencies. Residuals
      for the best-fit model are also shown in the bottom panel.}
    \label{fig:spinev}
\end{figure}

To determine the pulse frequency of the source we used the XRT window-timing
mode event data which has sufficient counting statistics and time resolution. After applying
the standard filtering criteria, we selected events with energies $0.3-10$\,keV
from the source-centered circle with radius of 25 pixels and applied barycentric
correction to the photon arrival times. To determine the spin frequency of the
source for each observation we performed a search for significant peaks around
the source pulse period using the H-test \citep{1989AA...221..180D}. To estimate the
uncertainty of the obtained value we used the same approach as
\cite{2002ApJ...575L..21M}, i.e. assumed that one sigma uncertainty for the most
significant frequency peak in periodogram corresponds to the drop $\Delta
Z^2=1$ with respect to the peak value. The results are presented in
Fig.~\ref{fig:spinev} and are consistent with values reported by {\it
Fermi}/GBM pulsar
project,\footnote{\url{https://gammaray.nsstc.nasa.gov/gbm/science/pulsars/lightc
urves /smcx3.html}} \citep{2017arXiv170102336T,2017arXiv170102983W}.

Note that modulation of the pulse frequency associated with the orbital motion
is apparent and has to be taken into the account when determining the intrinsic
spin frequency of the NS. The parameters of the binary orbit were
estimated for the current 2016--2017 outburst by \cite{2017arXiv170102336T} and
\cite{2017arXiv170102983W}. The
discrepancies, particularly in the obtained orbital period value, likely arise
due to the difficulties in modeling of complex intrinsic spin
evolution of the source.

To obtain an improved orbital solution we followed the approach similar to that
by \cite{2017arXiv170102336T}, i.e. estimated the intrinsic spin frequency of
the pulsar based on the expected accretion-induced spin-up. We note that
angular momentum transferred to the NS by the accretion disk is
proportional to the accretion rate and thus is expected to dominate the spin
evolution of the NS at high luminosities. The details of interaction
of the disk with the magnetosphere are not thus very important and it is
sufficient to consider only the accelerating torque (which, however, still
depends on the magnetosphere size) with the spin-up rate defined as
\citep{1982AZh....59..888L}
\be\label{eq0}
\dot{\nu}=\dot{M}\sqrt{GMR_{\rm d}}/2 \pi I,
\ee
where the inner disk radius $R_{\rm d}=k R_{\rm A}$ is assumed to constitute
some fraction of the Alfv\'enic radius, and $I$ is NS moment of inertia.
Note that this is essentially the same model as used by
\cite{2017arXiv170102336T}. The main difference is that we use the bolometric
light curve rather than flux in the soft band to estimate the accretion rate
$\dot{M}=R_{\rm NS}L_{\rm X}/GM$. Another difference is that we include also
the spin frequency measurements reported by {\it Fermi}/GBM into the fit to improve statistics.
Finally, \cite{2017arXiv170102336T} do not account for the model systematics
associated with the fact that uncertainties in the observed accretion rate
inevitably translate to a systematic uncertainty in the predicted frequency
when integrating over the outburst. On the other hand, this uncertainty can be
easily estimated directly from the dispersion of model predictions for a set of
light curves simulated based on the observed fluxes and uncertainties (once
initial estimate for $R_{\rm d}$ is obtained). We estimated the model systematics
to increase from zero at the beginning of the outburst to
$\sim2\times10^{-6}$\,Hz at the end of the outburst, and added it in quadrature
to the statistical uncertainties to obtain the final fit and estimate the
uncertainties for the orbital parameters presented in Table~\ref{tab:opar}.
Note that using the bolometric light curve allows to obtain a much better fit
than reported by \cite{2017arXiv170102336T} even without accounting for model
systematics.

\begin{table}
    \begin{center}
        \caption{Best-fit orbital parameters of \smcx3.}\label{tab:opar}
        \begin{small}
    \begin{tabular}{c c}
    \hline
    \hline
        Parameter & Value \\
        \hline
        Orbital period (d) & 45.07(5)\\
        $a_x \sin i$ (light seconds) & 189(1) \\
        $e$ & 0.231(6) \\
        $\omega$ (degrees)& 206(2) \\
        $T_\omega$ (MJD) & 57631.5(2) \\
        $\chi^2$/dof & 117.2/91\\
        \hline
    \end{tabular}
   \begin{center}{
    Note:  Parameters are based on the full sample of the {\it Fermi}/GBM and {\it Swift}/XRT data.
  Uncertainties are quoted at $1\sigma$ confidence level including the model systematics.
    The fit statistics for the same parameters without inclusion of the model systematics is $\chi^2=3992$ for 91 dof.
    }
\end{center}
\end{small}
\end{center}
\end{table}


\subsection{Pulse profile and pulsed fraction}
\label{sec:puls}

Excellent time and energy resolutions of the {\it NuSTAR} observatory
allowed us to investigate temporal properties of \smcx3\ at energies
above 20 keV in details for the first time. Moreover, two observations
performed on the rising and declining parts of the outburst gave us 
a possibility to trace evolution of these properties at different mass
accretion rates.

\begin{figure}
\includegraphics[width=0.96\columnwidth,bb=25 160 590 680]{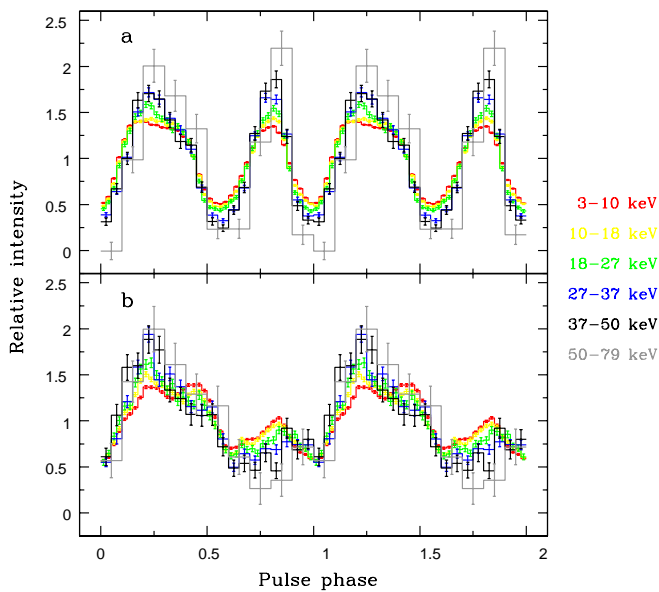}
\caption{Dependence of the pulse profile of \smcx3\ on energy in two {\it NuSTAR} observations. Different panels correspond to different X-ray luminosities: (a) $L_{\rm bol}=10.2\times10^{38}$ \lum (ObsID 90201035002), (b) $L_{\rm bol}=1.9\times10^{38}$ \lum\ (ObsID 90201041002). Different energy bands are shown with different colours (shown on the right). The profiles are normalized by the mean flux in each energy band and plotted twice for clarity.
}
\label{fig:pprof}
\end{figure}

The source pulse profiles in six energy bands from 3 to 79 keV are
shown in Fig.~\ref{fig:pprof} for two luminosities: $L_{\rm
  bol}=10.2\times10^{38}$ \lum\ (top panel; ObsID 90201035002) and
$L_{\rm bol}=1.9\times10^{38}$ \lum\ (bottom panel; ObsID
90201041002). To determine the pulse period and to produce the pulse
profile we used the combined light curves from both modules in order
to get the better statistics \citep[see][for the description of the
  procedure]{2015ApJ...809..140K}. To produce the pulse profile in
some energy range the corresponding light curve was folded with the
corrected for the orbital motion spin periods $P_{\rm s}=7.80954(5)$~s
and $P_{\rm s}=7.77084(7)$~s for first and second observations, respectively.

In both luminosity states the pulse profile exhibits a simple double-peak
shape with peaks separated by half phase. The main difference is that
the intensity of both peaks is approximately the same in the high
state, whereas in the observation with the lower luminosity the
intensity of the first (main) peak is significantly higher. The energy
dependence is very weak and noticeable mainly in the second
observation. Namely, the main peak (at phase $\sim0.25$) has a
double-peak substructure at low energies, gradually changing to a
single-peak with the energy increase. At the same time, the second
peak (at phase $\sim0.75$) is disappearing at higher energies.

\begin{figure}
\includegraphics[width=0.96\columnwidth,bb=15 270 515 680]{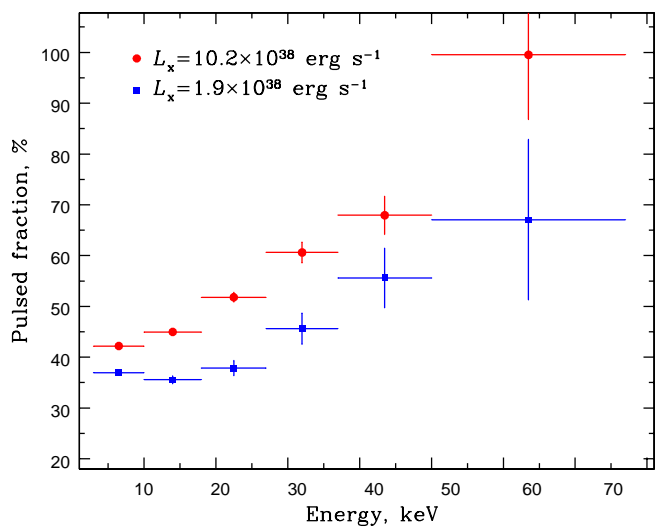}
\caption{Dependence of the pulsed fraction on energy in two {\it NuSTAR} observations with different X-ray luminosities: $L_{\rm bol}=10.2\times10^{38}$ \lum\ (red circles; ObsID 90201035002) and $L_{\rm bol}=1.9\times10^{38}$ \lum\ (blue squares; ObsID 90201041002).
}
\label{fig:pfrac}
\end{figure}

The pulsed fraction calculated using the standard
definition\footnote{$\mathrm{PF}=(F_\mathrm{max}-F_\mathrm{min})/(F_\mathrm{max}+F_\mathrm{min})$,
where $F_\mathrm{max}$ and $F_\mathrm{min}$ are maximum and minimum fluxes in
the pulse profile, respectively.} as a function of the energy is presented in
Fig.~\ref{fig:pfrac}. For both observations the pulsed fraction is
increasing towards higher energies, that is typical for the majority of X-ray
pulsars \citep{2009AstL...35..433L}. Note that during observation with the higher
luminosity the pulsed fraction was significantly higher throughout the energy range.

As can be seen from Figs. \ref{fig:pprof} and \ref{fig:pfrac} the
pulsed fraction is around 100\% at hard X-rays during the first
observation.  It is a very important fact since it rules out the
presence of a strong geometrical beaming of the emission by the
accretion disk, often used to artificially decrease the mass
accretion rate onto the NS in pulsing ULXs \citep[see e.g., ][]{2017MNRAS.468L..59K}. Another characteristics of
the pulsed fraction dependence on the energy is its smoothness. It is
consistent with the absence of any strong features (e.g., CSRF) in the
source energy spectrum (see Sect.~\ref{sec:spec}). Indeed, the
non-monotonic dependence of the pulsed fraction on energy around the
cyclotron line was shown to be typical for a few X-ray pulsars
\citep{2006MNRAS.371...19T,2009AstL...35..433L,2009A&A...498..825F}.

\subsection{Spectroscopy}
\label{sec:spec}

\begin{figure}
\centering
\includegraphics[width=0.98\columnwidth, bb=40 360 550 690]{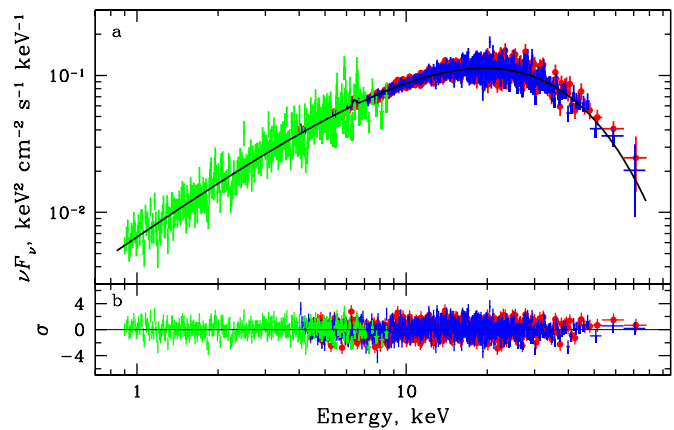}
\caption{ The broadband spectrum of \smcx3\ ({\it upper panel}) measured with {\it
    Swift}/XRT telescope (green points; ObsID 00034673045) and two {\it NuSTAR} modules
  FPMA and FPMB (red and blue points, respectively; ObsID 90201041002). Black solid line
  represents the best fit model consisting of exponentially cutoff
  power-law with addition of the fluorescent iron line.
  Corresponding residuals to the best-fit model are shown in the {\it lower panel}.  }\label{fig:spec}
\end{figure}

To supplement the {\it NuSTAR} data in the soft energy band and to determine
better the broad band spectrum of \smcx3\ we used the nearest in time {\it
Swift}/XRT observations (ObsIDs 00034673002 and 00034673045, respectively). The
broadband spectrum in the high luminosity state was described with the same model as
used by \citet{2016ATel.9404....1P} with following best-fit parameters values: the
photon index $\Gamma=0.89\pm0.01$, the folding energy $E_{\rm
fold}=14.88\pm0.14$ keV, the black body temperature $kT=0.89\pm0.05$ keV, the
iron line energy $6.38\pm0.03$ keV and width $0.37\pm0.05$ keV
(the equivalent width of $69\pm5$ eV). Note that the continuum parameters
are somewhat different from the ones reported by
\citet{2016ATel.9404....1P} due to the inclusion of the XRT data in
our analysis. The spectrum {in the low luminosity state} can be well
approximated by the exponentially cutoff power-law with photon index of
$\Gamma=0.62\pm0.01$ and folding energy of $E_{\rm fold}=14.16\pm0.17$ keV. A
fluorescent iron line at the energy of $6.40\pm0.13$ keV with the equivalent
width of $18\pm8$ eV is also tentatively registered in the spectrum. {Because
the statistic was not good enough to determine the line width directly, it
was fixed at the value $0.37$, determined for the first observations, that
roughly corresponded to the {\it NuSTAR} energy resolution
\citep{2013ApJ...770..103H}.} No thermal component is required for the second
observation. The corresponding broadband spectrum of the source is presented
in Fig.~\ref{fig:spec}.

To take into account the uncertainty in the instrument calibrations
as well as the lack of strict simultaneity of observations by {\it NuSTAR} and {\it
Swift}, cross-calibration constants between them were included in the
spectral modeling.

No other obvious spectral features (primarily, the CSRF) were found in both
{\it NuSTAR} spectra of \smcx3. To quantify this conclusion we used an approach, initially
proposed by \citet{2005AstL...31...88T} and recently improved by
\citet{2017AstL...43..175S}. The spectral models were modified by addition of
the {\sc gabs} component from the {\sc xspec} package. The CSRF energy
$E_{\rm cyc}$ was varied within the 6--62 keV energy range with the step of 3
keV. A corresponding line width was varied within the 4--8 keV range with the
step of 2 keV. Each combination of the lines position and width were fixed
and the resulting model was used to approximate the source spectrum. As a
result, none of such combinations resulted in a significant improvement of
the fit and only the upper limits for the optical depth of $\sim 0.25$ and
$\sim 0.19$ ($3\sigma$) can be obtained for the first and second
{\it NuSTAR} observations, respectively.

\begin{figure}
\centering
\includegraphics[width=0.97\columnwidth,bb=55 250 460 680]{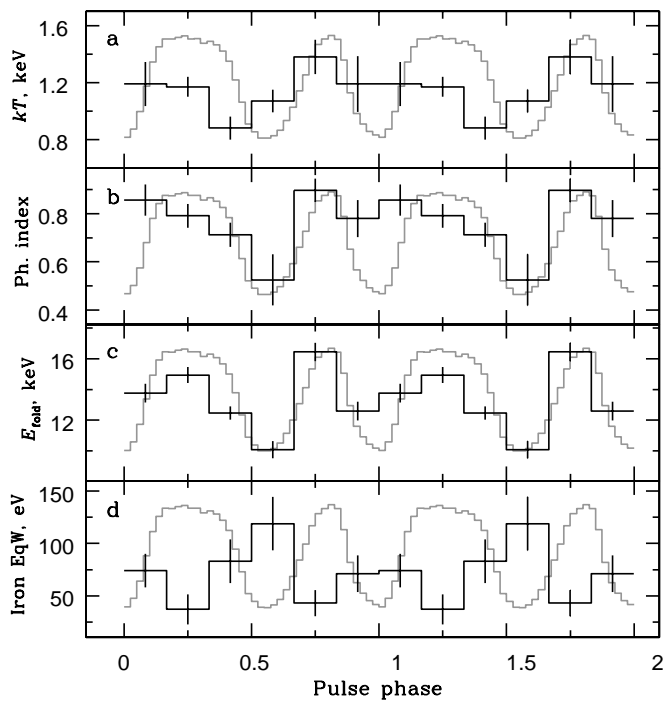}
\caption{Variations of spectral parameters of \smcx3\ over the pulse
  phase for the high luminosity state ({\it NuSTAR} observation ID
  90201035002). The black histogram in the panels represents: (a)
  black body temperature, (b) photon index, (c) folding energy and (d)
  equivalent width of the iron emission line. The grey line in each
  panel shows the pulse profile in a wide energy
  range.}\label{fig:pph}
\end{figure}

Finally, we performed also a pulse phase-resolved spectroscopy of the
emission of \smcx3\ using the {\it NuSTAR} data. It was found that the
source spectrum at different phases is well described with the same
models as were used for the average spectra. An example of the 
source's spectral parameters variations over the pulse is shown in 
Fig.\,\ref{fig:pph} for the high luminosity state. It is clearly seen
that all continuum parameters are slightly varying with the pulse phase 
in the ranges of $\Gamma \sim 0.5-0.9$, $E_{\rm fold} \sim 11-16$ keV, 
$kT \sim 0.9-1.3$ keV. It is interesting to note that the equivalent width
of the iron emission line is also variable with the phase, demonstrating 
a possible anti-correlation with the pulse intensity. Such a behavior 
was earlier reported for several other sources 
\citep[see, e.g.][]{2009essu.confE..70T,2017AstL...43..175S} and can be used 
for the tomography of the matter distribution in the vicinity of the NS. For the low luminosity state 
the behavior of parameters is quite similar: the photon index and
folding energy are varying with the pulse phase in the ranges
of $\Gamma \sim 0.5-0.7$ and $E_{\rm fold} \sim 11-17$ keV,
respectively. The emission iron line is detected significantly only 
in one phase bin. Again, no indications for the CSRF
were found in these spectra.

\section{Discussion}
\label{sec:dis}

\subsection{Critical and maximal luminosities}
\label{sec:crit}

In the case of very strong magnetic field  Compton scattering cross
section is strongly reduced \citep[see, e.g.,
][]{2016PhRvD..93j5003M}. This reduces an impact of the radiative
pressure and was shown to be required to allow the accreting X-ray
pulsar to exhibit super-Eddington luminosities during giant outbursts
similar to one observed from \smcx3. Particularly, it is possible due
to the rise of the accretion column above the NS surface
\citep{1976MNRAS.175..395B,2015MNRAS.447.1847M}.  Critical luminosity
$L^*$ dividing two regimes of accretion is a function of the
magnetic field strength in the vicinity of the NS surface. Therefore,
observational constraint of $L^*$ can be used to obtain independent
estimates of the magnetic field. It is worth to note here that
observational evidence of a transition through the critical luminosity
was recently found in the classical XRP V~0332+53
\citep{2017MNRAS.466.2143D}.

It is interesting that \smcx3\ exhibits a substantial change of its
pulse profile between MJD $\sim57680$ and MJD $\sim57700$ when its
main peak shifted by $\sim\pi/2$ \citep[see Fig. 4
  in][]{2017arXiv170102983W}. We argue that this change happend when
the bolometric luminosity of the source was $\sim
(2-3)\times10^{38}$~\lum\ is caused by the disappearance of the
accretion column and corresponding modification of the intrinsic X-ray
beaming from the pulsar
\citep[][]{1973A&A....25..233G}. Interpretation of this luminosity as
a critical one results in the estimates of the magnetic field of
$B\simeq (2-3)\times 10^{13}\,{\rm G}$ \citep{2015MNRAS.447.1847M}.

Maximal accretion luminosity of the NS depends on the magnetic field
strength as well \citep{2015MNRAS.454.2539M}. The bolometric peak
luminosity of \smcx3\ achieved during the 2016--2017 outburst is
$\sim2.5\times10^{39}$~\lum\ exceeding the Eddington limit by an order
of magnitude. According to the accretion column theory by
\cite{2015MNRAS.454.2539M} so high luminosity is impossible if the NS
magnetic field is less than $\sim 2\times10^{13}\,{\rm G}$. This value
agrees well with the one derived above from the critical
luminosity. Both methods refer to the magnetic field strength in the
region of the main energy release, i.e. in immediate vicinity of the
NS surface.

\subsection{Accretion torque}

Evolution of the intrinsic spin frequency of accreting pulsars is
driven by angular momentum transfer from the accretion disk, and
possibly by braking mechanisms associated with interaction of the
magnetosphere with the disk \citep{1979ApJ...234..296G,
  Wang87,1995MNRAS.275..244L,2016ApJ...822...33P}. In case of \smcx3,
however, the accelerating torque is expected to dominate, and indeed,
as already mentioned in Section~\ref{sec:opars}, no evidence for
braking has been observed.  The observed spin evolution is thus fully
consistent with pure spin-up by the disk which is trivial to calculate
and allows to estimate the magnetic field.  As follows from
Eq.~(\ref{eq0}), the magnetosphere size or, equivalently, magnetic
field strength of the NS is one of the model parameters that
can be obtained during the fit of the orbital parameters if physical
parameters of the NS and relation of the inner disk and Alfv\'enic
radii are known or assumed.  For standard NS with mass and
radius of $1.4M_\odot$ and 10\,km respectively, and coupling factor
$k=0.5$ the field strength is well constrained at
$B=2.55(1)\times10^{12}$\,G. This is consistent with the value
reported by \citet{2014MNRAS.437.3863K} based on the spin evolution
between outbursts, and an order of magnitude lower than estimated
above based on the observed critical and maximal luminosities.

\subsection{Propeller effect}
\label{sec:prop}

The distinctive property of accretion onto the highly magnetized NS
is a strong centrifugal barrier produced by the rotating magnetosphere. This
barrier does not allow the accreting matter to penetrate into the magnetosphere
if the velocity of the field lines is higher than corresponding local Keplerian
velocity, that is known as propeller effect
\citep{1975A&A....39..185I,1986ApJ...308..669S}. In other words the accretion
is only possible if the magnetospheric radius $R_{\rm m}$ does not exceed the
co-rotation radius $R_{\rm c}$.

\begin{figure}
\centering
\includegraphics[width=0.98\columnwidth, bb=20 285 550 675]{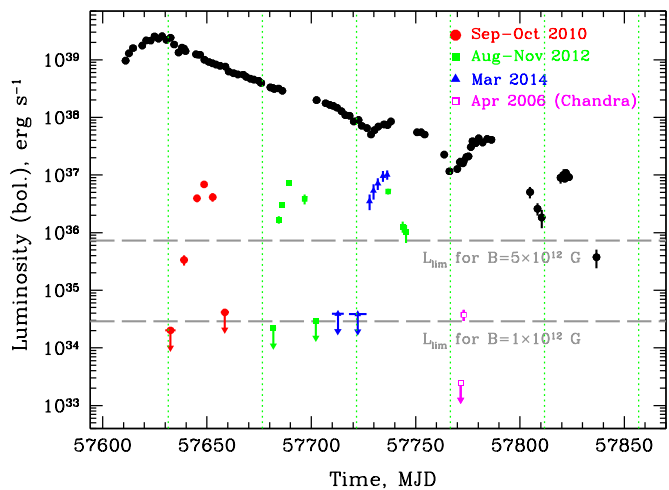}
\caption{The bolometric light curve of \smcx3\ as seen by the {\it
    Swift}/XRT telescope obtained during giant outburst in 2016--2017,
  type I outbursts in September-October 2010, August-November 2012 and
  March 2014 are shown with black circles, red circles, green squares
  and blue triangles, respectively. {\it Chandra} measurement from
  April 2006 is represented with magenta square. The $3\sigma$ upper
  limits are shown with arrows of the corresponding color. The
  historical data were shifted by an integer number of the orbital periods
  obtained in the current work (see Table~\ref{tab:opar}) for
  illustrative purpose. The luminosity is calculated from the unabsorbed flux under
  assumption of the distance to the source $d=62$ kpc and bolometric
  correction factors from Fig.~\ref{fig:bolom}. The horizontal dashed
  lines show the upper and lower limits for the threshold luminosity
  for the propeller regime onset (see the text).  Green vertical
  dotted lines correspond to the times of the periastron
  passages. }\label{fig:hist}
\end{figure}

Because the magnetospheric radius depends on the mass accretion rate, we can link the
transition luminosity $L_{\rm lim}$ with the spin period of the NS
and its magnetic field strength. The corresponding equation can be derived by
the equating the magnetospheric radius to the co-rotation radius
\citep[e.g.,][]{2002ApJ...580..389C}:
\be\label{eq1}
L_{\rm lim}(R) \simeq \frac{GM\dot{M}_{\rm lim}}{R}
\simeq 4 \times 10^{37} k^{7/2} B_{12}^2 P^{-7/3} M_{1.4}^{-2/3} R_6^5 \,\textrm{erg s$^{-1}$} ,
\ee
where $P$ is the NS spin period in seconds, $B_{12}$ is the strength of the dipole component of the
magnetic field in units of $10^{12}$~G, $M_{1.4}$ and $R_6$ are the NS
mass and radius in units of 1.4M$_\odot$ and $10^6$~cm, respectively. A factor
$k$ relates the magnetospheric radius to the classical Alfv\'en radius in the
case of disk accretion and is usually taken $k=0.5$ \citep{GL1978}.

On 2017 February 1, \smcx3\ was still very
bright to expect the transition to the propeller regime for the NS
with standard magnetic field. Therefore, we investigated the archival
{\it Swift}/XRT and {\it Chandra} data. The archival {\it Swift}
observations were analysed as described above. The source spectra
available as part of the {\it Chandra} Source
Catalog\footnote{\url{http://cxc.harvard.edu/csc/};
  \citet{2010ApJS..189...37E}} were fit using the {\sc xspec} and the
same model as for the {\it Swift}/XRT spectra. The resulting light
curves based on all available data as well as $3\sigma$ upper limits
are shown in Fig.~\ref{fig:hist}.

From Fig.~\ref{fig:hist} one can see that \smcx3\ was significantly
detected at low luminosities during several type I outbursts. The faintest
state of the source with bolometric luminosity of
$\sim3.5\times10^{34}$\,\lum\ was observed by the {\it Chandra}
observatory on 2006, April 26. Just one day before that the source was
not detected with $3\sigma$ upper limit of
$\sim2.5\times10^{33}$\,\lum. The lowest significant luminosity of
\smcx3\ can serve as an estimate for the threshold of the propeller
regime onset $L_{\rm lim}\sim3\times10^{34}$~\lum. On the other hand,
based only on the {\it Swift}/XRT data the limiting luminosity can be
as high as $L_{\rm lim}\sim7\times10^{35}$\,\lum. Therefore, we use
here very conservative range of luminosities from
$\sim3\times10^{34}$~\lum\ to $\sim7\times10^{35}$\,\lum\ to estimate
the dipole component of the magnetic field (shown by horizontal dashed
lines in Fig.~\ref{fig:hist}).
Substituting the measured spin period of \smcx3\ ($P=7.81$ s) and the
range of limiting propeller luminosity $L_{\rm lim}$ to the
Eq.~(\ref{eq1}) and assuming the standard mass and radius of the
NS we can estimate the dipole component of the NS magnetic
field as $B\sim(1-5)\times10^{12}$~G.

Interestingly, both methods (accretion torque and the propeller effect)
measuring the dipole component of the magnetic field well agree with each
other and give significantly lower strength in comparison to the
methods sensitive to the magnetic field in the vicinity of the NS
(critical and maximal luminosities). This fact can point to the strongly non-dipole
configuration of the NS magnetic field with multipoles by an order of
magnitude stronger than dipole component within the emission region.
Worth mentioning that absence of the CSRF is consistent with the above picture.
Another possibility to resolve this discrepancy is significant deviation
of the parameter $k$ from the standard value of 0.5 adopted in our
study. Namely, the estimate of the dipole component of the magnetic
field $B_{\rm dipole}$ will coincide with multipole component
$B_{\rm multipole}$ for $k\approx0.15$.

\begin{figure}
\centering
\includegraphics[width=0.98\columnwidth,bb=60 275 565 680]{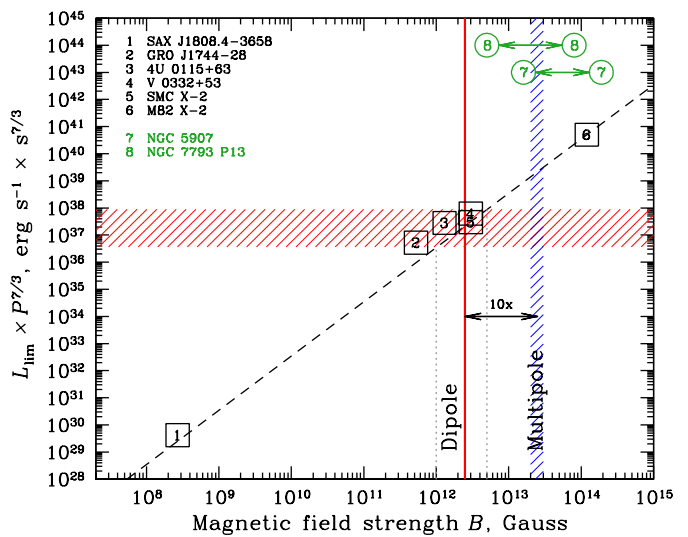}
\caption{Correlation between a combination of the propeller limiting
  luminosity and the pulsar spin period, $L_{\rm lim}P^{7/3}$, and
  independently determined magnetic field strength $B$ for six
  pulsating sources (black squares; adopted from
  \cite{2016A&A...593A..16T}). Dashed line represents the theoretical
  dependence given by Eq.~(\ref{eq1}) for $k=0.5$.  Estimate of the
  limiting luminosity and corresponding dipole field strength for
  \smcx3 are shown with horizontal shaded region and dotted vertical
  lines, respectively. The solid red vertical line indicates the
  dipole field estimate from the observed spin evolution of the
  source. Constraints on the field strength in vicinity of the NS
  based on the observed critical and maximal luminosities are shown
  with vertical shaded region. Note an order of magnitude discrepancy
  between the field estimates at the magnetosphere and in vicinity of
  the NS. Green circles show similar discrepancy between two field
  components in the pulsating ULXs in NGC 5907 and NGC 7793 P13.  }
\label{fig:collect}
\end{figure}

However, this assumption seems to be inconsistent with the results obtained
previously for other sources.
Indeed, up to date the propeller effect was observed in SAX\,J1808.4-3658
\citep{2008ApJ...684L..99C}, GRO\,J1744-28
\citep{1997ApJ...482L.163C}, 4U\,0115+63 \citep{2016A&A...593A..16T},
V\,0332+63 \citep{2016A&A...593A..16T}, SMC\,X-2
\citep{2017ApJ...834..209L}, and the accreting magnetar M82 X-2
\citep{2016MNRAS.457.1101T}. All these sources are collected in
Fig.~\ref{fig:collect} where the combination of the propeller limiting
luminosity and corresponding pulse period, $L_{\rm lim}P^{7/3}$, are
compared with magnetic field strength measured independently
\citep[mainly based on the observed CSRF energy, see ][and references therein]{2016A&A...593A..16T}.
Note that for all six sources the limiting
luminosity seems to agree with the theoretical dependence given by
Eq.~(\ref{eq1}) for $k=0.5$ shown with dashed line.
Discrepancy between two field components was also pointed out in two pulsating ULXs in NGC 5907 \citep{2017Sci...355..817I} and NGC 7793 P13 \citep{2017MNRAS.466L..48I}. Interestingly, in both sources multipole components are estimated to be an order of magnitude stronger in comparison to the dipole component (see green circles in Fig.~\ref{fig:collect}), similarly to the case of \smcx3.

\section{Conclusion}
\label{sec:con}

In this paper we report the estimate of the magnetic field strength in
a bright X-ray pulsar \smcx3 which can be considered the closest
ULX-pulsar. The source exhibited a giant outburst in July 2016 --
March 2017 with the peak bolometric luminosity of
$\sim2.5\times10^{39}$~\lum. The entire outburst had been monitored
with the {\it Swift}/XRT and {\it Fermi}/GBM telescopes, as well as
with the {\it NuSTAR} observatory. The collected data allowed us to
estimate the magnetic field strength of the NS in \smcx3\ using
several independent methods based on the spectral and timing
properties of X-ray emission from the system.

The dipole component of the magnetic field was determined using the
accretion torque models and observation of the transition to the
propeller regime at the limiting luminosity in the range of
$\sim(0.3-7)\times10^{35}$~erg~s$^{-1}$ both resulting in relatively
weak strength of about $(1-5)\times10^{12}$~G. On the other hand,
there is evidence for a much stronger field strength in the immediate
vicinity of the NS surface. In particular the transition from super-
to sub-critical accretion regime associated with cease of an accretion
column and very high peak luminosity favor the magnetic field of
$\sim(2-3)\times10^{13}$~G. Absence of the CSRF
in the broadband X-ray spectrum of the source obtained with {\it
  NuSTAR} is also consistent with this estimate.

This discrepancy makes \smcx3\ a good candidate for a NS with strong multipole
configuration of the magnetic field, and an intermediate source between
classical X-ray pulsars and accreting magnetars which may constitute an
appreciable fraction of ULX population. Alternatively, this discrepancy
can be resolved if one assumes that the accretion disk in \smcx3 pushes much
deeper into magnetosphere than normally expected. However, the reason for such
behavior would be unclear as it was not observed in other objects with wide
range of luminosities for which similar analysis was conducted.

On the other hand, existence of higher field multipoles was also claimed for
another pulsating ULXs \citep{2015MNRAS.448L..40E,2017Sci...355..817I,2017MNRAS.466L..48I}. One can thus speculate
that a complex structure of the NS magnetic field may thus be a common feature
for this class of objects.

\begin{acknowledgements}
This work was supported by the Russian Science Foundation grant 14-12-01287
(SST, AAL, AAM), the Foundations' Professor Pool, the Finnish Cultural Foundation and the Academy of Finland grant 268740  (JP).  We also
acknowledge the support from the COST Action MP1304.
\end{acknowledgements}

\bibliographystyle{aa}
\bibliography{allbib}
\end{document}